\documentclass[%
 reprint,
superscriptaddress,
reprint,
 amsmath,amssymb,
 aps,
pra,
floatfix,
]{revtex4-2}

\usepackage{graphicx}
\usepackage{dcolumn}
\usepackage{bm}
\usepackage{hyperref}

\usepackage[dvipsnames]{xcolor}
\usepackage{soul}
\usepackage{ulem}

\begin{document}


\title{A Landau-Zener formula for the Adiabatic Gauge Potential}

\author{Gabriel Cardoso}\affiliation{Tsung-Dao Lee Institute,
Shanghai Jiao Tong University, Shanghai, 201210, China}
\email{gabriel.jg.cardoso@outlook.com}

\date{\today}
\begin{abstract} 

By the adiabatic theorem, the probability of non-adiabatic transitions in a time-dependent quantum system vanishes in the adiabatic limit. The Landau-Zener (LZ) formula gives the leading functional behaviour of the probability close to this limit. On the other hand, in counterdiabatic dynamics one achieves effectively adiabatic evolution at finite driving speed by adding an extra field which suppresses non-adiabatic transitions: the adiabatic gauge potential (AGP). We investigate the mechanism by which the AGP suppresses the transition probability, changing it from the LZ formula to exactly zero. Quantitatively, we find that adding the AGP to the Hamiltonian modifies the LZ formula by a universal prefactor, independent of the adiabatic parameter, which vanishes in the counterdiabatic regime. Qualitatively, this prefactor can be understood as arising from the Aharonov-Bohm phases generated by the AGP between different paths in the complex time plane. Finally, we show that these results extend to a class of integrable time-dependent quantum Hamiltonians by proving that the AGP preserves their integrability condition.

\end{abstract}

\maketitle


The adiabatic theorem states that a system prepared in an eigenstate of a time-dependent Hamiltonian will remain in the corresponding instantaneous eigenstate, provided that the Hamiltonian changes slowly compared to the timescale set by the gap to the other states \cite{adiabatictheorem}. Motivated by a variety of applications in quantum technology, recent years have seen a lot of interest in control protocols where effectively adiabatic evolution can be realized even when the Hamiltonian changes rapidly \cite{quantumgate,adquantumcomputing,annealingpra17,geometric,openquantum}. In counterdiabatic driving, the idea is to modify the Hamiltonian by an extra term, which is engineered so as to minimize the probability of non-adiabatic transitions \cite{berry,campostaprl13,shortcuts,genshortcut17,reviewrvm19}. The term which gives total suppression of the transitions is known as the adiabatic gauge potential (AGP), since it is the gauge generator of the unitary rotation into the time-dependent eigenbasis \cite{agppolkovnikov}. Here we address the qualitative and quantitative mechanism by which the AGP suppresses non-adiabatic transitions.

More specifically, the probability of non-adiabatic transitions approaches zero in a characteristic way in the adiabatic limit. In the Landau-Zener-St\"uckelberg-Majorana (LZSM) model \cite{landau,zener,majorana,stueckelberg1932theory,SHEVCHENKO20101}, the probability vanishes exponentially with the adiabatic parameter. Although this result, the Landau-Zener formula (LZ), was derived for a specific model, a generalization was later given by Dykhne \cite{dykhne} to non-degenerate two-level systems. The exponent is fixed by the position of the degeneracy point of the analytically extended eigenvalues in the complex time plane which lies closest to the real axis. Davis, Hwang and Pechukas \cite{davis,pechukas} formalized a method for deriving the Dykhne formula as well as its possible generalizations when the eigenvalues have different analytic structures, known as the Dykhne-Davis-Pechukas (DDP) method. In \cite{berrygeometric,joyegeometric}, Berry and Joye independently showed that the geometric phase can gain an imaginary part when extended to the complex time plane, and adds a prefactor to the Dykhne formula independent of the adiabatic parameter $\delta$. Sub-leading corrections to the exponent can be found as a power series on the adiabatic parameter \cite{joyeexpansion}.

It is interesting to investigate how the AGP changes this formula. For this purpose, we compute the LZ estimate for a one-parameter family of Hamiltonians which interpolates between the LZSM model and its counterdiabatic correction, where the interpolating parameter is the strength of the AGP. As expected, the result interpolates between the LZ formula and exactly zero as a function of this parameter. We find that the Landau-Zener exponent, coming from the dynamical phase, is not modified. Instead, the presence of the AGP leads to a universal prefactor independent of the adiabatic parameter. This prefactor is topological,  in the sense that it arises from relative phases for different paths in the complex time plane appearing due to a singularity of the AGP, much like in the Aharonov-Bohm mechanism \cite{aharonov1959significance}. At the end, we extend our conclusions to a class of integrable time-dependent Hamiltonians \cite{integrable}, by proving that the characteristic flatness condition in these systems is preserved by the AGP.

\section{Transition probability}

The problem we consider is the following. Let $H(t)$ be a non-degenerate, time-dependent, two-level Hamiltonian with a gap of order $\mathcal{E}$, and which changes over a time-scale $T$. It is then interesting to rewrite the time-dependent Schr\"odinger equation (TDSE) in the dimensionless form
\begin{equation}
    i\delta\partial_\tau|\psi\rangle=h|\psi\rangle,\label{eq:TDSE}
\end{equation}
where $\tau=t/T$ is the the dimensionless ``slow time", $h(\tau)=\frac{1}{\mathcal{E}}H$ and
\begin{equation}
    \delta=\frac{\hbar}{\mathcal{E}T}
\end{equation}
is the adiabatic parameter. The instantaneous eigenvalues $e_n$ and eigenstates $|n\rangle$ are given by
\begin{align}
    h(\tau)|n(\tau)\rangle=e_n(\tau)|n(\tau)\rangle\label{eq:eigenstates},\,\, n=0,1.
\end{align}
We consider the solution of (\ref{eq:TDSE}) in the form
\begin{equation}
    |\psi(\tau)\rangle=\sum_nc_n(\tau)e^{-\frac{i}{\delta}\int_0^\tau e_n-i\int_0^\tau a_n}|n(\tau)\rangle,\label{eq:eigendecomp}
\end{equation}
which starts from the ground state,
\begin{equation}
    \lim_{\tau\to-\infty} c_n(\tau)=\delta_{n,0}.\label{eq:initial}
\end{equation}
Here we also defined the Berry connection,
\begin{equation}
    a_n=-i\langle n|\partial_\tau|n\rangle.
\end{equation}

Then the probability of non-adiabatic transitions is
\begin{equation}
    P\equiv \lim_{\tau\to+\infty}|c_1(\tau)|^2.
\end{equation}
One can compute $P$ for Hamiltonians with different parameters. Then the adiabatic theorem says that its dependence on the adiabatic parameter, $P(\delta)$, is such that
\begin{equation}
    P\xrightarrow[\delta\to 0]{}0,
\end{equation}
and the LZ formula gives the leading behaviour of $P(\delta)$ in approaching this limit
\begin{equation}
    P\sim e^{-\pi/\delta}.\label{eq:LZformula}
\end{equation}
This formula is quite generic, though it is known that it can be modified if the Hamiltonian depends on additional parameters independent of $\delta$ \cite{abanovrequist}.

On the other hand, suppose the eigenstates (\ref{eq:eigenstates}) depend on time through the parameter $\theta$ and let $U(\theta)$ be the corresponding unitary operator which diagonalizes the Hamiltonian. Then the adiabatic gauge potential (AGP) is
\begin{equation}
    \mathcal{A}_\theta=-i\delta U\partial_\theta U^\dagger,
\end{equation}
and it is easy to check by direct computation that the original eigenstates (\ref{eq:eigenstates}),
\begin{equation}
    e^{-\frac{i}{\delta}\int_0^\tau e_n-i\int_0^\tau \dot{\theta}a_n}|n(\tau)\rangle,
\end{equation}
where $\theta=\theta(\tau)$, are exact solutions of the TDSE with the new Hamiltonian
\begin{equation}
    h(\theta)+\dot{\theta}\mathcal{A}_\theta.
\end{equation}
In other words, adding the AGP to the Hamiltonian replaces (\ref{eq:LZformula}) by
\begin{equation}
    P=0.
\end{equation}
We consider the interpolation between these two results by calculating $P_\eta(\delta)$, the probability of non-adiabatic transitions for the one-parameter family of Hamiltonians
\begin{equation}
    h+\eta \dot{\theta}\mathcal{A}_\theta,\label{eq:heta}
\end{equation}
where the strength $\eta$ of the AGP term tunes the Hamiltonian from the LZSM model at $\eta=0$ to the counterdiabatic regime at $\eta=1$. We find that, for small $\delta$,
\begin{equation}
    P_\eta\sim \cos^2\frac{\eta\pi}{2}e^{-2\eta/3}e^{-\pi/\delta}.\label{eq:modLZ1}
\end{equation}
Interestingly, the usual LZ exponent of order $1/\delta$ is unchanged. Instead, the AGP suppresses this estimate by a delta-independent prefactor which exactly vanishes for $\eta\to 1$. We show that this prefactor appears due to a topological phase introduced by the AGP, which is singular in the complex time plane. This is different, for example, from the mechanism in \cite{takayoshi2021nonadiabatic}, where the counterdiabatic regime appears because the dynamical exponent changes.

\section{The DDP method}

Note that formulas (\ref{eq:LZformula},\ref{eq:modLZ1}) vanish in all orders of perturbation theory in the parameter $\delta$. To capture the non-perturbative result, we proceed by a generalization of the Dykhne-Davis-Pechukas (DDP) method, which makes use of the analytic properties of the Hamiltonian. Let us first review the more familiar LZ case (for more details, see \cite{pechukas}). The LZSM hamiltonian is
\begin{equation}
    H=
        \begin{pmatrix}
           a t  & b \\
            b & -a t
        \end{pmatrix}.
    \label{eq:landauzener}
\end{equation}
The eigenvalues $\pm\sqrt{a^2t^2+b^2}$ have an avoided crossing at $t=0$, with minimal gap $2b$. Thus it is natural to take $\mathcal{E}=b$ and $T=b/a$, so that the TDSE takes the form (\ref{eq:TDSE}) with
\begin{align}
    h=e(\tau)[\sin\theta\sigma^x+\cos\theta\sigma^z],\,\,\delta=\frac{\hbar a}{b^2},\label{eq:LZH}
\end{align}
where
\begin{equation}
     e(\tau)=\sqrt{1+\tau^2},\,\,\tan\theta=\frac{1}{\tau}.\label{eq:etauthetatau}
\end{equation}
The instantaneous eigenstates are
\begin{align}
    |0\rangle=\begin{pmatrix}
    -\sin\frac{\theta}{2}\\
    \cos\frac{\theta}{2}
    \end{pmatrix}, \,\, |1\rangle=\begin{pmatrix}
    \cos\frac{\theta}{2}\label{eq:lzstates}\\
    \sin\frac{\theta}{2}
    \end{pmatrix},\,\, e_{0,1}=\mp e(\tau),
\end{align}
and we consider the decomposition of the solution in terms of the eigenstates as in (\ref{eq:eigendecomp}). Then the coefficients $c_n(\tau)$ satisfy the equations
\begin{align}
    \dot{c}_0&=p_{01}e^{\frac{i}{\delta}\int_0^\tau (e_0-e_1)  +i\int_0^\tau(a_0-a_1)}c_1(\tau),\label{eq:ceq1}\\
    \dot{c}_1&=p_{10}e^{\frac{i}{\delta}\int_0^\tau (e_1-e_0)  +i\int_0^\tau(a_1-a_0)}c_0(\tau),\label{eq:ceq2}
\end{align}
with the anti-hermitian coefficients $p_{nm}=-\langle n|\partial_\tau|m\rangle$. We consider the solution with initial condition (\ref{eq:initial}).

\begin{figure}
    \centering
    \includegraphics[width=0.4\textwidth]{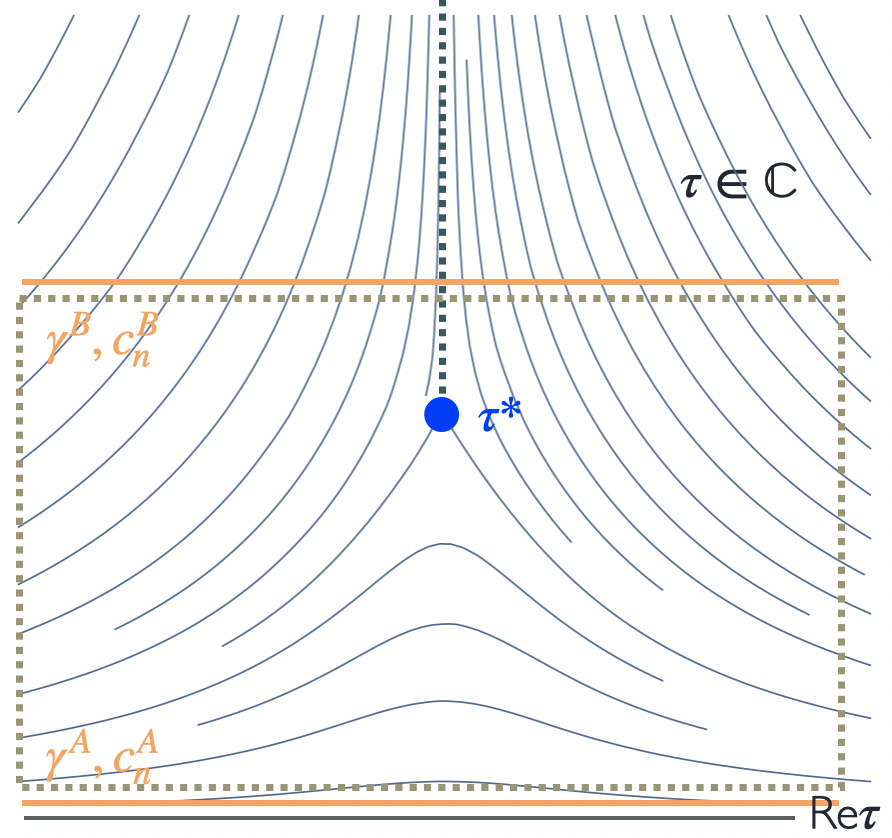}
    \caption{Level lines of the function $\Delta^A(\tau)$ (\ref{eq:deltacondition}) for the eigenvalues of the LZSM model. There is a square-root branch point $\tau^*$, from which an extra level line extends vertically. We place the branch cut on top of this extra level line. Then one can see that both the real axis $\gamma^A$ and the curve $\gamma^B$ satisfy the condition of non-decreasing $\Delta(\tau)$ for the adiabatic theorem.}
    \label{fig:LZ}
\end{figure}

The essence of the method lies in extending these equations to complex $\tau$. Since the Hamiltonian (\ref{eq:LZH}) is analytic in the complex plane, the solution of the TDSE (\ref{eq:TDSE}) is well-defined everywhere\footnote{The argument generalizes to the case where the Hamiltonian extends analytically to only a stripe around the real axis and containing the closes degeneracy point of the eigenstates, but the distinction does not play a role here.}. However, a subtlety appears when writing the equations for the amplitudes $c_n$ (\ref{eq:ceq1},\ref{eq:ceq2}): although the solution $|\psi(\tau)\rangle$ is single-valued, the eigenstates are not. Indeed, the eigenvalues $\pm e(\tau)=\pm\sqrt{1+\tau^2}$ have square-root branch points at the complex degeneracies $\pm\tau^*$, where
\begin{equation}
    \tau^*=i,
\end{equation}
so that going around one of these points amounts to swapping the two eigenstates. Therefore, the decomposition (\ref{eq:eigendecomp}) of the solution of the TDSE only makes sense with respect to a specific curve, which we will keep track of by using superscripts: $c_n^{A,B,...}$ will denote the amplitudes of $|\psi(\tau)\rangle$ in terms of the eigenstates defined continuously along the curve $\gamma^{A,B,...}$. Additionally, since we are ultimately interested in the amplitudes $c_n(\tau)$ for real $\tau$, we introduce branch cuts from $\pm\tau^*$ to infinity for when extending the eigenstates continuously from the real axis as in figure \ref{fig:LZ}. With this definition, the labelling of adiabatic amplitudes $c_n^B$ on a given curve $\gamma^B$ coincides with the one on the real axis $\gamma^A$ if the curve does not cross the branch cut, and gets relatively swapped if the curve crosses it. The LZ formula follows from comparing the amplitudes corresponding to different curves in the adiabatic limit.

In the limit $\delta\to 0$, the equations (\ref{eq:ceq1},\ref{eq:ceq2}) are dominated by the dynamical phase factors, and this leads to an important result. Let $\gamma^A(s)$ be a curve on the complex plane such that $\text{Re}[\gamma^A(s)]\xrightarrow[s\to\pm\infty]{}\pm\infty$. If, on $\gamma^A$, the function
\begin{equation}
    \Delta^A(\tau)=\text{Im}[\int_0^{\tau=\gamma^A(s)} (e_0^A-e_1^A)ds']\label{eq:deltacondition}
\end{equation}
is non-increasing, then one can show that the adiabatic theorem is valid, that is, 
\begin{equation}
    c_n(\gamma^A(s))\xrightarrow[s\to +\infty]{\delta\to0}\delta_{n0}.
\end{equation}
This result includes the adiabatic theorem, since $\Delta(\tau)$ vanishes on the real axis $\gamma^A$, but the point is that there are other curves on which it also applies, thus earning it the name of adiabatic theorem in the complex plane \cite{pechukas}. In figure \ref{fig:LZ}, we plot the level lines of the function $\Delta(\tau)$. It increases away from the real axis, and the level line splits into three at the branch point $\tau^*$. Consider now the curve $\gamma^B$. To the left of the branch cut, it traverses the decreasing level lines, while to the right it traverses the increasing level lines. But, since the labelling of eigenstates is swapped as the curve crosses the branch cut, this means that the condition (\ref{eq:deltacondition}) is also satisfied on all of $\gamma^B$.

We compare the eigenstate decompositions of $|\psi(\tau)\rangle$ on the real axis $\gamma^A$ and on the curve $\gamma^B$. Since to the left of the cut the labels of the eigenstates agree, the initial condition becomes
\begin{align}
    c_n^A(\tau)\xrightarrow[\tau\to-\infty]{}\delta_{n0},\label{eq:adcla}\\
    c_n^B(\tau)\xrightarrow[\tau\to-\infty]{}\delta_{n0}.\label{eq:adclb}
\end{align}
And, since in both $\gamma^A$ and $\gamma^B$ the adiabatic theorem is satisfied, we have that
\begin{align}
    c_n^A(\tau)\xrightarrow[\tau\to+\infty]{\delta\to 0}\delta_{n0},\label{eq:adca}\\
    c_n^B(\tau)\xrightarrow[\tau\to+\infty]{\delta\to 0}\delta_{n0}.\label{eq:adcb}
\end{align}
However, due to the relative crossing of the branch point, the adiabatic limit answer (\ref{eq:adcb}) gives the leading correction to the adiabatic limit in (\ref{eq:adca}) by matching the coefficients of the wavefunction. With a slight abuse of notation,
\begin{align}
    |\psi(+\infty)\rangle &=e^{-i\int_0^{\gamma^B(\infty)}\left(\frac{e_0}{\delta}-a_0\right)}|0^B(+\infty)\rangle\\
    &=\sum_n c_n^A(+\infty)e^{-i\int_0^{\gamma^A(\infty)}\left(\frac{e_n}{\delta}-a_n\right)}|n^A(+\infty)\rangle.\nonumber
\end{align}
Since $|0^B(+\infty)\rangle=|1^A(+\infty)\rangle$, matching this coefficient gives
\begin{align}
    c_1^A(+\infty)&=e^{i\int_0^{\gamma^A(\infty)}\left(\frac{e^A_1}{\delta}-a^A_1\right)}e^{-i\int_0^{\gamma^B(\infty)}\left(\frac{e^B_0}{\delta}-a^B_0\right)}\\
    &=e^{-\frac{i}{\delta}\int_0^{\tau^*}(e_0-e_1)},
\end{align}
where in the second line we deformed the integration contour, represented by the dotted line in figure \ref{fig:LZ} \footnote{The line integrals along the curve $\gamma^B$, $\int_0^{\tau=\gamma^B(s)}$ are defined to be on the path: from zero, down the real axis to $-\infty$, then to $\gamma(-\infty)$ and then along $\gamma^B$ up to $\gamma^B(s)$. This definition (as a limit) makes sense provided that the Hamiltonian $h(\tau)$ smoothly approaches well-defined limits $h_{\pm}$ as $\text{Re}[\tau]\to\pm\infty$ \cite{pechukas}.}. Because the Landau-Zener Hamiltonian (\ref{eq:LZH}) is purely real, the Berry phase factor vanishes. The result is Dykhne's formula: the probability of transitions is fixed by the phase integral of the gap between the eigenstates up to the closest branch point,
\begin{equation}
    P=|c_1^A(+\infty)|^2\sim e^{-\frac{2}{\delta}\text{Im}[\int_0^{\tau^*}(e_1-e_0)]}.\label{eq:dykhne}
\end{equation}
Using the explicit eigenvalues of the LZSM model we find
\begin{equation}
    P\sim e^{-\frac{4}{\delta}\text{Im}[\int_0^{\tau^*}\sqrt{1+\tau^2}]}=e^{-\pi/\delta},
\end{equation}
which is the LZ formula.

\section{The AGP phase}

Let us now see how this calculation is changed by the AGP. The LZSM hamiltonian (\ref{eq:LZH}) is diagonalized by
\begin{equation}
    U(\theta)=\begin{pmatrix}
        -\sin\frac{\theta}{2}&\cos\frac{\theta}{2}\\
        \cos\frac{\theta}{2}&\sin\frac{\theta}{2}
    \end{pmatrix}=e^{-i\frac{\theta}{2}\sigma^y}\sigma^x,
\end{equation}
with time-dependence $\theta=\theta(\tau)$ from (\ref{eq:etauthetatau}), so that the AGP takes on the form
\begin{equation}
    \dot{\theta}\mathcal{A}_\theta=-\frac{\delta}{2(1+\tau^2)}\sigma^y.\label{eq:AGP}
\end{equation}
Thus the Hamiltonian (\ref{eq:heta}), parametrized by $\eta\in[0,1]$, is given by
\begin{align}
    h(\eta)&=e(\tau)[\sin\theta\sigma^x+\cos\theta\sigma^z]+\eta\dot{\theta}\mathcal{A}_\theta\\
    &=\sigma^x+\tau\sigma^z-\frac{\eta\delta}{2(1+\tau^2)}\sigma^y.\label{eq:hnew}
\end{align}
We see that, while the LZSM Hamiltonian is analytic everywhere, the presence of the AGP leads to poles at $\pm\tau^X$, where
\begin{equation}
    \tau^X=i.
\end{equation}
In particular, this means that the solution of the TDSE (\ref{eq:TDSE}) is not well-defined everywhere. In fact, in the punctured plane $\mathbb{C}\setminus \{\tau^X,-\tau^X\}$, it is multi-valued.

\begin{figure}
    \centering
    \includegraphics[width=0.4\textwidth]{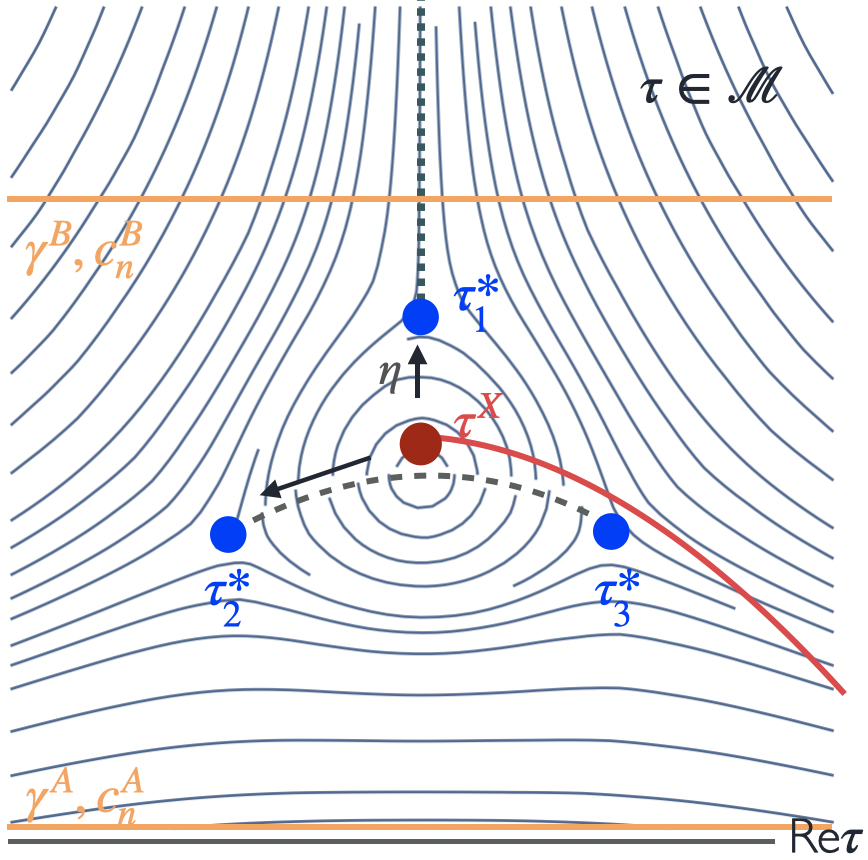}
    \caption{Level lines of the function $\Delta^A(\tau)$ (\ref{eq:deltacondition}) for the eigenvalues of the counterdiabatic LZSM model (\ref{eq:heta}). We introduce a Dirac string extending from the singularity of the Hamiltonian at $\tau^X$ (red). In the upper half of the complement of the Dirac string, $\mathcal{M}^+$, there are three branch points, from which we define the branch cuts as above: one extending up from $\tau_1^*$ along the vertical level line, and one connecting $\tau_2^*$ to $\tau_3^*$. Then one can see that both the real axis $\gamma^A$ and the curve $\gamma^B$ satisfy the condition of non-decreasing $\Delta(\tau)$ for the adiabatic theorem.}
    \label{fig:AGP}
\end{figure}

A simple work-around for dealing with singular gauge fields that have non-trivial holonomy is known from the theory of Dirac's magnetic monopole \cite{dirac1931quantised}: we make the domain of the TDSE simply connected by introducing a Dirac string extending from each pole to infinity (represented by the red line in figure \ref{fig:AGP}). Then, on the complement of the Dirac string, $\mathcal{M}$, the solution is well-defined and single-valued. For simplicity, we take the strings to become parallel to, and very close to the real axis as $\text{Re}[\tau]\to +\infty$.

On the other hand, the eigenvalues of the new Hamiltonian (\ref{eq:hnew}) are
\begin{equation}
    e_{0,1}=\mp\frac{1}{1+\tau^2}\sqrt{(1+\tau^2)^3+\frac{\eta^2\delta^2}{4}},\label{eq:eigenAGP}
\end{equation}
which vanish at the six square-root branch points $\pm\tau_{1,2,3}^*$ at
\begin{equation}
    \tau_{1,2,3}^*=i\left[1+\left(\frac{\eta\delta}{2}\right)^{\frac{2}{3}}e^{i\frac{2\pi}{3}k}\right]^{\frac{1}{2}},
\end{equation}
with $k=0,1,2$. Compare figures \ref{fig:LZ} and \ref{fig:AGP}. For $\eta> 0$, there are three branch points $\tau_{1,2,3}^*$ surrounding each pole of the Hamiltonian $\tau^X$, while for $\eta\to 0$ all of these collapse to the single branch point of the LZSM model.

Again, we find that the expansion (\ref{eq:eigendecomp}) of the wavefunction is not well-defined in the whole $\mathcal{M}$. Thus we extend the labelling of eigenstates from the real time axis to the upper half-plane sector $\mathcal{M}^+$ by introducing branch cuts as in figure \ref{fig:AGP}: one cut extending from $\tau_1^*$ to infinity along the imaginary axis and one connecting $\tau_2^*$ to $\tau_3^*$. The level lines of $\Delta(\tau)$ are plotted in figure \ref{fig:AGP}. Outside the region occupied by the singular points, they look similar to the LZSM case, so that one can easily check that the adiabatic theorem applies to the curve $\gamma^B$, and we have equations (\ref{eq:adcla}-\ref{eq:adcb}). However, since the curves $\gamma^A$ and $\gamma^B$ are separated by the Dirac string, the two expansions cannot be immediately matched at $\text{Re}[\tau]\to +\infty$. Indeed, the wavefunction itself is not continuous upon jumping across the string, and the discontinuity is found by integrating the TSDE on a path going around the Dirac string. Recalling that we take the string to lie arbitrarily close to the real axis, we find the matching condition
\begin{equation}
    |\psi(+\infty)\rangle=e^{-\eta\frac{i}{\delta}\oint_{\tau^X}d\tau\dot{\theta}\mathcal{A}_\theta}|\psi(+\infty+i\epsilon)\rangle,\label{eq:psimatch}
\end{equation}
where the exponent corresponds to a residue integral of the AGP around the pole at $\tau^X$,
\begin{equation}
    e^{-\frac{i}{\delta}\eta\oint\mathcal{A}_\theta d\theta}=e^{i\eta\frac{\pi}{2}\sigma^y},\label{eq:nonabelianphase}
\end{equation}
so that
\begin{align}
    e^{-\frac{i}{\delta}\oint_{\tau^X}d\theta\mathcal{A}_\theta}|0^B(+\infty)\rangle=&\cos\left(\frac{\eta\pi}{2}\right)|1^A(+\infty)\rangle\\
    &-\sin\left(\frac{\eta\pi}{2}\right)|0^A(+\infty)\rangle.
\end{align}
Therefore, comparing the $|1^A(+\infty)\rangle$ coefficients of (\ref{eq:psimatch}), we find
\begin{align}
    c_1^A(+\infty)&=e^{i\int_0^{\gamma^A(\infty)}\left(\frac{e^A_1}{\delta}-a^A_1\right)}e^{-i\int_0^{\gamma^B(\infty)}\left(\frac{e^B_0}{\delta}-a^B_0\right)}\cos\frac{\eta\pi}{2}\nonumber\\
    &=\cos\frac{\eta\pi}{2}e^{-\frac{i}{\delta}\int_0^{\tau_1^*}(e_0-e_1)}e^{i\int_0^{\tau_1^*}(a_0-a_1)},
\end{align}
where the integrations from zero to the branch point $\tau_1^*$ in these expressions are along paths circumventing the branch cuts. Thus the transition probability is
\begin{equation}
    P_\eta\sim \cos^2\frac{\eta\pi}{2}e^{2\text{Im}[\int_0^{\tau_1^*}(a_1-a_0)]}e^{-\frac{2}{\delta}\text{Im}[\int_0^{\tau_1^*}(e_1-e_0)]}.\label{eq:modLZ}
\end{equation}
We see that there are two modifications from the Dykhne formula (\ref{eq:dykhne}): the Berry phase factor, which in this case is non-zero; and the first prefactor, which appears due to the topological phase introduced by the AGP.

\begin{figure}
    \centering
    \includegraphics[width=0.45\textwidth]{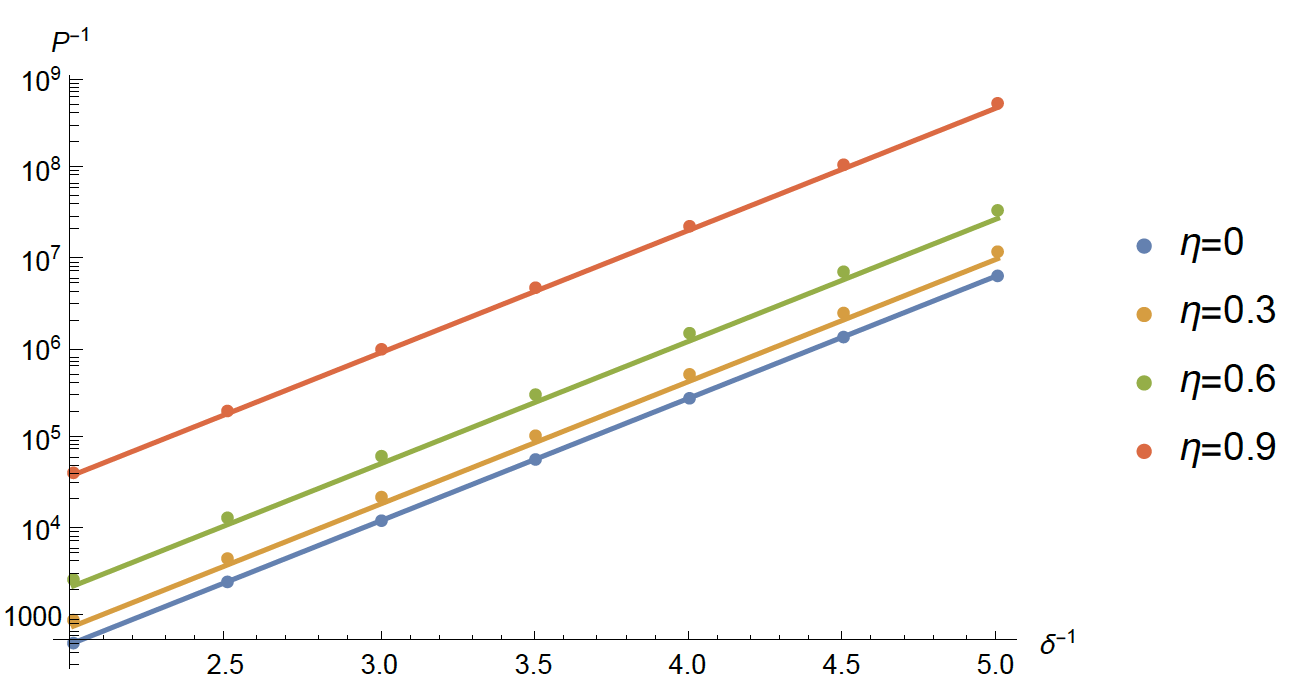}
    \caption{Inverse probability of non-adiabatic transitions $P^{-1}$ as a function of inverse adiabatic parameter $\delta^{-1}$ for Hamiltonians (\ref{eq:heta}) with different values of $\eta$. The solid lines are the formula (\ref{eq:modLZ}), and the dots are the results of numerically solving the TDSE (\ref{eq:TDSE}) for the transition probability $P=|c_1(200)|^2$ with initial condition $c_n(-200)=\delta_{n0}$.}
    \label{fig:probabilities}
\end{figure}

Again, the integrals can be evaluated by deforming the contour, though one should be careful with the branch cuts. In appendix \ref{sec:appexponent}, we find that the $O(1/\delta)$ and $O(1)$ terms in the dynamical phase are
\begin{equation}
    e^{-\frac{2}{\delta}\text{Im}[\int_0^{\tau_1^*}(e_1-e_0)]}=e^{-\pi/\delta+2\eta/3},
\end{equation}
while the geometric phase factor, although not identically vanishing, doesn't contribute at this order. Finally, we find the modified LZ formula
\begin{equation}
    P_\eta\sim \cos^2\frac{\eta\pi}{2}e^{-2\eta/3}e^{-\pi/\delta}.\label{eq:modLZ2}
\end{equation}
As shown in figure \ref{fig:probabilities}, this formula agrees with the results found by numerically solving the TDSE for different values of $\eta$.

Thus we find that the AGP leads to a new prefactor in the LZ formula, which is responsible for the total suppression of non-adiabatic transitions for $\eta\to 1$. The change in the transition probabilities appears due to the non-trivial holonomy (\ref{eq:nonabelianphase}) of the adiabatic gauge potential, which generates relative phases between the different paths in the complex time plane. In this way we arrive at the interesting conclusion that the AGP suppresses non-adiabatic transitions by Aharonov-Bohm interference in complex time.

\section{Integrable time-dependent Hamiltonians}

Formula (\ref{eq:modLZ}) is particular to the LZSM model, but as an application we now show that this intuitive picture about the AGP mechanism applies to a wider class of integrable time-dependent quantum Hamiltonians \cite{integrable}.

In these systems, integrability refers to the following method for finding the exact solution of the TDSE,
\begin{equation}
    i\partial_t|\psi\rangle =H_0|\psi\rangle,
\end{equation}
in the $(-\infty,\infty)$ time interval: there are extra parameters $x^j$ (for example, couplings) in the Hamiltonian $H_0$ and corresponding Hamiltonian operators $H_j$ such that the collection $(H_{\mu})=(H_0,H_j)$ satisfies the compatibility condition
\begin{equation}
    \partial_{\mu}H_{\nu}-\partial_{\nu}H_{\mu}+i[H_{\mu},H_{\nu}]=0,\label{eq:flatH}
\end{equation}
where $(x^{\mu})=(t,x^j)$. Then one can extend the TDSE to an equation for parallel transport on the whole $(x^{\mu})$ space with connection form $H_{\mu}dx^{\mu}$,
\begin{equation}
    i\partial_{\mu}|\psi\rangle=H_{\mu}|\psi\rangle.\label{eq:integrableTDSE}
\end{equation}
Since equation (\ref{eq:flatH}) is the zero-curvature condition for this connection, it implies that the solution of (\ref{eq:integrableTDSE}) is independent of the path of integration.

For a given initial condition at $t\to -\infty$ and fixed $x^j$, one can use this freedom to find the solution for $t\to +\infty$ by solving the TDSE on a convenient path in the $(x^\mu)$ space. It is interesting to choose a path for which $|x|$ is always large, because generally this guarantees that the eigenvalues are far apart and the solution only gains the appropriate phases multiplying each instantaneous eigenstate. The exception is when the path intersects the hypersurfaces where an avoided crossing happens. Crossing one of these hypersurfaces is equivalent to a Landau-Zener problem for the corresponding states which come close together. This reduces the matrix of transition probabilities to a product of Landau-Zener factors
\begin{equation}
    \begin{pmatrix}
       1-P & P\\
       P & 1-P
    \end{pmatrix},
    \label{eq:lzmatrix}
\end{equation}
where the LZ transition probabilities are given by (\ref{eq:LZformula}) with the appropriate adiabatic parameter for each crossing hypersurface. One can think that the LZSM model is the building block of these time-dependent integrable quantum Hamiltonians. Examples which fall under this paradigm include multi-level Landau-Zener models \cite{countermultilz,multilz,gridlz,malla}, Gaudin magnets and driven BCS Hamiltonians \cite{gaudin}.

In these examples, the Hamiltonians $H_{\mu}$ are real and symmetric. One can then separate the real and immaginary parts of (\ref{eq:flatH}),
\begin{align}
    \partial_{\mu} H_{\nu}-\partial_{\nu} H_{\mu} = 0,\label{eq:derivativesflat}\\
    [H_{\mu},H_{\nu}]=0\label{eq:commutatorflat}.
\end{align}
We argue that the way in which the probabilities of non-adiabatic transitions are suppressed by the AGP in these models is by altering the LZ factors (\ref{eq:lzmatrix}) in the way we discussed above. The reason is that the AGP preserves the integrability condition, which means that the argument leading to the localization of the non-adiabatic transitions to the crossing hypersurfaces can be repeated in the presence of the AGP.

More precisely, besides the AGP for the original Hamiltonian,
\begin{equation}
    \mathcal{A}_0=-iU\partial_\tau U^\dagger,
\end{equation}
one can define the potential $\mathcal{A}_\mu$ corresponding to counterdiabatic evolution in the direction $x^\mu$. We show in appendix \ref{sec:appintegrable} that the corrected Hamiltonians also satisfy the flatness condition,
\begin{equation}
    \partial_{\mu}(H_{\nu}+\mathcal{A}_{\nu})-\partial_{\nu}(H_{\mu}+\mathcal{A}_{\mu})+i[H_{\mu}+\mathcal{A}_{\mu},H_{\nu}+\mathcal{A}_{\nu}]=0,
\end{equation}
so that the extended parameter space picture still holds. One can use the same contour which led to the factorization into LZ factors (\ref{eq:lzmatrix}). It is easy to see that the AGP is suppressed by the gap everywhere except on the crossing hypersurfaces, where it takes the form (\ref{eq:AGP}) with the corresponding adiabatic parameter for the reduced LZ problem.

\section{Discussion}

The adiabatic gauge potential completely suppresses non-adiabatic transitions. In particular, it kills off the Landau-Zener estimate of the transition probability. We investigated the mechanism by which this happens, by using the extension of the TDSE to complex time. As in the Landau-Zener case, the leading contributions come from paths which go around the branch point of the eingenvalues which lies closest to the real axis. However, the non-trivial topology of the adiabatic gauge potential leads to phase factors which eventually cancel these contributions in the counterdiabatic regime. Then, by giving a proof of the compatibility of the AGP with the flatness condition of integrable time-dependent quantum Hamiltonians, we showed that there is a larger class of models where this intuition about the AGP mechanism is valid. In fact, it is possible that these conclusions extend beyond the integrable case to more general models with analytic (or meromorphic) Hamiltonians. It would also be interesting to understand how the cosine prefactor derived here is related to the generic formulas derived in \cite{berry1993universal} by the method of superadiabatic renormalization. We leave these questions for future investigation.

\begin{acknowledgements}
This work was partially supported by NSF under Grant NSF DMR-2116767 and by the Simons Center for Geometry and Physics. The author is grateful to Alexander G Abanov for many helpful discussions.
\end{acknowledgements}

\bibliography{Refs}

\begin{thebibliography}{36}%
\makeatletter
\providecommand \@ifxundefined [1]{%
 \@ifx{#1\undefined}
}%
\providecommand \@ifnum [1]{%
 \ifnum #1\expandafter \@firstoftwo
 \else \expandafter \@secondoftwo
 \fi
}%
\providecommand \@ifx [1]{%
 \ifx #1\expandafter \@firstoftwo
 \else \expandafter \@secondoftwo
 \fi
}%
\providecommand \natexlab [1]{#1}%
\providecommand \enquote  [1]{``#1''}%
\providecommand \bibnamefont  [1]{#1}%
\providecommand \bibfnamefont [1]{#1}%
\providecommand \citenamefont [1]{#1}%
\providecommand \href@noop [0]{\@secondoftwo}%
\providecommand \href [0]{\begingroup \@sanitize@url \@href}%
\providecommand \@href[1]{\@@startlink{#1}\@@href}%
\providecommand \@@href[1]{\endgroup#1\@@endlink}%
\providecommand \@sanitize@url [0]{\catcode `\\12\catcode `\$12\catcode
  `\&12\catcode `\#12\catcode `\^12\catcode `\_12\catcode `\%12\relax}%
\providecommand \@@startlink[1]{}%
\providecommand \@@endlink[0]{}%
\providecommand \url  [0]{\begingroup\@sanitize@url \@url }%
\providecommand \@url [1]{\endgroup\@href {#1}{\urlprefix }}%
\providecommand \urlprefix  [0]{URL }%
\providecommand \Eprint [0]{\href }%
\providecommand \doibase [0]{https://doi.org/}%
\providecommand \selectlanguage [0]{\@gobble}%
\providecommand \bibinfo  [0]{\@secondoftwo}%
\providecommand \bibfield  [0]{\@secondoftwo}%
\providecommand \translation [1]{[#1]}%
\providecommand \BibitemOpen [0]{}%
\providecommand \bibitemStop [0]{}%
\providecommand \bibitemNoStop [0]{.\EOS\space}%
\providecommand \EOS [0]{\spacefactor3000\relax}%
\providecommand \BibitemShut  [1]{\csname bibitem#1\endcsname}%
\let\auto@bib@innerbib\@empty
\bibitem [{\citenamefont {Born}\ and\ \citenamefont
  {Fock}(1928)}]{adiabatictheorem}%
  \BibitemOpen
  \bibfield  {author} {\bibinfo {author} {\bibfnamefont {M.}~\bibnamefont
  {Born}}\ and\ \bibinfo {author} {\bibfnamefont {V.}~\bibnamefont {Fock}},\
  }\bibfield  {title} {\bibinfo {title} {Beweis des adiabatensatzes},\
  }\href@noop {} {\bibfield  {journal} {\bibinfo  {journal} {Zeitschrift
  f{\"u}r Physik}\ }\textbf {\bibinfo {volume} {51}},\ \bibinfo {pages} {165}
  (\bibinfo {year} {1928})}\BibitemShut {NoStop}%
\bibitem [{\citenamefont {Santos}\ \emph {et~al.}(2016)\citenamefont {Santos},
  \citenamefont {Silva},\ and\ \citenamefont {Sarandy}}]{quantumgate}%
  \BibitemOpen
  \bibfield  {author} {\bibinfo {author} {\bibfnamefont {A.~C.}\ \bibnamefont
  {Santos}}, \bibinfo {author} {\bibfnamefont {R.~D.}\ \bibnamefont {Silva}},\
  and\ \bibinfo {author} {\bibfnamefont {M.~S.}\ \bibnamefont {Sarandy}},\
  }\bibfield  {title} {\bibinfo {title} {{Shortcut to adiabatic gate
  teleportation}},\ }\href {https://doi.org/10.1103/physreva.93.012311}
  {\bibfield  {journal} {\bibinfo  {journal} {Physical Review A}\ }\textbf
  {\bibinfo {volume} {93}},\ \bibinfo {pages} {012311} (\bibinfo {year}
  {2016})}\BibitemShut {NoStop}%
\bibitem [{\citenamefont {Hegade}\ \emph {et~al.}(2021)\citenamefont {Hegade},
  \citenamefont {Paul}, \citenamefont {Ding}, \citenamefont {Sanz},
  \citenamefont {Albarrán-Arriagada}, \citenamefont {Solano},\ and\
  \citenamefont {Chen}}]{adquantumcomputing}%
  \BibitemOpen
  \bibfield  {author} {\bibinfo {author} {\bibfnamefont {N.~N.}\ \bibnamefont
  {Hegade}}, \bibinfo {author} {\bibfnamefont {K.}~\bibnamefont {Paul}},
  \bibinfo {author} {\bibfnamefont {Y.}~\bibnamefont {Ding}}, \bibinfo {author}
  {\bibfnamefont {M.}~\bibnamefont {Sanz}}, \bibinfo {author} {\bibfnamefont
  {F.}~\bibnamefont {Albarrán-Arriagada}}, \bibinfo {author} {\bibfnamefont
  {E.}~\bibnamefont {Solano}},\ and\ \bibinfo {author} {\bibfnamefont
  {X.}~\bibnamefont {Chen}},\ }\bibfield  {title} {\bibinfo {title} {{Shortcuts
  to Adiabaticity in Digitized Adiabatic Quantum Computing}},\ }\href
  {https://doi.org/10.1103/physrevapplied.15.024038} {\bibfield  {journal}
  {\bibinfo  {journal} {Physical Review Applied}\ }\textbf {\bibinfo {volume}
  {15}},\ \bibinfo {pages} {024038} (\bibinfo {year} {2021})},\ \Eprint
  {https://arxiv.org/abs/2009.03539} {2009.03539} \BibitemShut {NoStop}%
\bibitem [{\citenamefont {Takahashi}(2017)}]{annealingpra17}%
  \BibitemOpen
  \bibfield  {author} {\bibinfo {author} {\bibfnamefont {K.}~\bibnamefont
  {Takahashi}},\ }\bibfield  {title} {\bibinfo {title} {{Shortcuts to
  adiabaticity for quantum annealing}},\ }\href
  {https://doi.org/10.1103/physreva.95.012309} {\bibfield  {journal} {\bibinfo
  {journal} {Physical Review A}\ }\textbf {\bibinfo {volume} {95}},\ \bibinfo
  {pages} {012309} (\bibinfo {year} {2017})}\BibitemShut {NoStop}%
\bibitem [{\citenamefont {Du}\ \emph {et~al.}(2019)\citenamefont {Du},
  \citenamefont {Liang}, \citenamefont {Yan},\ and\ \citenamefont
  {Zhu}}]{geometric}%
  \BibitemOpen
  \bibfield  {author} {\bibinfo {author} {\bibfnamefont {Y.}~\bibnamefont
  {Du}}, \bibinfo {author} {\bibfnamefont {Z.}~\bibnamefont {Liang}}, \bibinfo
  {author} {\bibfnamefont {H.}~\bibnamefont {Yan}},\ and\ \bibinfo {author}
  {\bibfnamefont {S.}~\bibnamefont {Zhu}},\ }\bibfield  {title} {\bibinfo
  {title} {{Geometric Quantum Computation with Shortcuts to Adiabaticity}},\
  }\href {https://doi.org/10.1002/qute.201900013} {\bibfield  {journal}
  {\bibinfo  {journal} {Advanced Quantum Technologies}\ }\textbf {\bibinfo
  {volume} {2}},\ \bibinfo {pages} {1900013} (\bibinfo {year}
  {2019})}\BibitemShut {NoStop}%
\bibitem [{\citenamefont {Yin}\ \emph {et~al.}(2022)\citenamefont {Yin},
  \citenamefont {Li}, \citenamefont {Allcock}, \citenamefont {Zheng},
  \citenamefont {Gu}, \citenamefont {Dai}, \citenamefont {Zhang},\ and\
  \citenamefont {An}}]{openquantum}%
  \BibitemOpen
  \bibfield  {author} {\bibinfo {author} {\bibfnamefont {Z.}~\bibnamefont
  {Yin}}, \bibinfo {author} {\bibfnamefont {C.}~\bibnamefont {Li}}, \bibinfo
  {author} {\bibfnamefont {J.}~\bibnamefont {Allcock}}, \bibinfo {author}
  {\bibfnamefont {Y.}~\bibnamefont {Zheng}}, \bibinfo {author} {\bibfnamefont
  {X.}~\bibnamefont {Gu}}, \bibinfo {author} {\bibfnamefont {M.}~\bibnamefont
  {Dai}}, \bibinfo {author} {\bibfnamefont {S.}~\bibnamefont {Zhang}},\ and\
  \bibinfo {author} {\bibfnamefont {S.}~\bibnamefont {An}},\ }\bibfield
  {title} {\bibinfo {title} {{Shortcuts to adiabaticity for open systems in
  circuit quantum electrodynamics}},\ }\href
  {https://doi.org/10.1038/s41467-021-27900-6} {\bibfield  {journal} {\bibinfo
  {journal} {Nature Communications}\ }\textbf {\bibinfo {volume} {13}},\
  \bibinfo {pages} {188} (\bibinfo {year} {2022})}\BibitemShut {NoStop}%
\bibitem [{\citenamefont {Berry}(2009)}]{berry}%
  \BibitemOpen
  \bibfield  {author} {\bibinfo {author} {\bibfnamefont {M.~V.}\ \bibnamefont
  {Berry}},\ }\bibfield  {title} {\bibinfo {title} {{Transitionless quantum
  driving}},\ }\href {https://doi.org/10.1088/1751-8113/42/36/365303}
  {\bibfield  {journal} {\bibinfo  {journal} {Journal of Physics A:
  Mathematical and Theoretical}\ }\textbf {\bibinfo {volume} {42}},\ \bibinfo
  {pages} {365303} (\bibinfo {year} {2009})}\BibitemShut {NoStop}%
\bibitem [{\citenamefont {Campo}(2013)}]{campostaprl13}%
  \BibitemOpen
  \bibfield  {author} {\bibinfo {author} {\bibfnamefont {A.~d.}\ \bibnamefont
  {Campo}},\ }\bibfield  {title} {\bibinfo {title} {{Shortcuts to adiabaticity
  by counterdiabatic driving.}},\ }\href
  {https://doi.org/10.1103/physrevlett.111.100502} {\bibfield  {journal}
  {\bibinfo  {journal} {Physical review letters}\ }\textbf {\bibinfo {volume}
  {111}},\ \bibinfo {pages} {100502} (\bibinfo {year} {2013})}\BibitemShut
  {NoStop}%
\bibitem [{\citenamefont {Torrontegui}\ \emph {et~al.}(2012)\citenamefont
  {Torrontegui}, \citenamefont {Ibáñez}, \citenamefont {Martínez-Garaot},
  \citenamefont {Modugno}, \citenamefont {Campo}, \citenamefont
  {Guéry-Odelin}, \citenamefont {Ruschhaupt}, \citenamefont {Chen},\ and\
  \citenamefont {Muga}}]{shortcuts}%
  \BibitemOpen
  \bibfield  {author} {\bibinfo {author} {\bibfnamefont {E.}~\bibnamefont
  {Torrontegui}}, \bibinfo {author} {\bibfnamefont {S.}~\bibnamefont
  {Ibáñez}}, \bibinfo {author} {\bibfnamefont {S.}~\bibnamefont
  {Martínez-Garaot}}, \bibinfo {author} {\bibfnamefont {M.}~\bibnamefont
  {Modugno}}, \bibinfo {author} {\bibfnamefont {A.~d.}\ \bibnamefont {Campo}},
  \bibinfo {author} {\bibfnamefont {D.}~\bibnamefont {Guéry-Odelin}}, \bibinfo
  {author} {\bibfnamefont {A.}~\bibnamefont {Ruschhaupt}}, \bibinfo {author}
  {\bibfnamefont {X.}~\bibnamefont {Chen}},\ and\ \bibinfo {author}
  {\bibfnamefont {J.~G.}\ \bibnamefont {Muga}},\ }\bibfield  {title} {\bibinfo
  {title} {{Shortcuts to adiabaticity}},\ }\href
  {https://doi.org/10.1016/b978-0-12-408090-4.00002-5} {\bibfield  {journal}
  {\bibinfo  {journal} {arXiv}\ }\textbf {\bibinfo {volume} {62}},\ \bibinfo
  {pages} {117} (\bibinfo {year} {2012})},\ \Eprint
  {https://arxiv.org/abs/1212.6343} {1212.6343} \BibitemShut {NoStop}%
\bibitem [{\citenamefont {Patra}\ and\ \citenamefont
  {Jarzynski}(2017)}]{genshortcut17}%
  \BibitemOpen
  \bibfield  {author} {\bibinfo {author} {\bibfnamefont {A.}~\bibnamefont
  {Patra}}\ and\ \bibinfo {author} {\bibfnamefont {C.}~\bibnamefont
  {Jarzynski}},\ }\bibfield  {title} {\bibinfo {title} {{Shortcuts to
  adiabaticity using flow fields}},\ }\href
  {https://doi.org/10.1088/1367-2630/aa924c} {\bibfield  {journal} {\bibinfo
  {journal} {New Journal of Physics}\ }\textbf {\bibinfo {volume} {19}},\
  \bibinfo {pages} {125009} (\bibinfo {year} {2017})}\BibitemShut {NoStop}%
\bibitem [{\citenamefont {Guéry-Odelin}\ \emph {et~al.}(2019)\citenamefont
  {Guéry-Odelin}, \citenamefont {Ruschhaupt}, \citenamefont {Kiely},
  \citenamefont {Torrontegui}, \citenamefont {Martínez-Garaot},\ and\
  \citenamefont {Muga}}]{reviewrvm19}%
  \BibitemOpen
  \bibfield  {author} {\bibinfo {author} {\bibfnamefont {D.}~\bibnamefont
  {Guéry-Odelin}}, \bibinfo {author} {\bibfnamefont {A.}~\bibnamefont
  {Ruschhaupt}}, \bibinfo {author} {\bibfnamefont {A.}~\bibnamefont {Kiely}},
  \bibinfo {author} {\bibfnamefont {E.}~\bibnamefont {Torrontegui}}, \bibinfo
  {author} {\bibfnamefont {S.}~\bibnamefont {Martínez-Garaot}},\ and\ \bibinfo
  {author} {\bibfnamefont {J.~G.}\ \bibnamefont {Muga}},\ }\bibfield  {title}
  {\bibinfo {title} {{Shortcuts to adiabaticity: Concepts, methods, and
  applications}},\ }\href {https://doi.org/10.1103/revmodphys.91.045001}
  {\bibfield  {journal} {\bibinfo  {journal} {Reviews of Modern Physics}\
  }\textbf {\bibinfo {volume} {91}},\ \bibinfo {pages} {045001} (\bibinfo
  {year} {2019})},\ \Eprint {https://arxiv.org/abs/1904.08448} {1904.08448}
  \BibitemShut {NoStop}%
\bibitem [{\citenamefont {Kolodrubetz}\ \emph {et~al.}(2017)\citenamefont
  {Kolodrubetz}, \citenamefont {Sels}, \citenamefont {Mehta},\ and\
  \citenamefont {Polkovnikov}}]{agppolkovnikov}%
  \BibitemOpen
  \bibfield  {author} {\bibinfo {author} {\bibfnamefont {M.}~\bibnamefont
  {Kolodrubetz}}, \bibinfo {author} {\bibfnamefont {D.}~\bibnamefont {Sels}},
  \bibinfo {author} {\bibfnamefont {P.}~\bibnamefont {Mehta}},\ and\ \bibinfo
  {author} {\bibfnamefont {A.}~\bibnamefont {Polkovnikov}},\ }\bibfield
  {title} {\bibinfo {title} {Geometry and non-adiabatic response in quantum and
  classical systems},\ }\href
  {https://doi.org/https://doi.org/10.1016/j.physrep.2017.07.001} {\bibfield
  {journal} {\bibinfo  {journal} {Physics Reports}\ }\textbf {\bibinfo {volume}
  {697}},\ \bibinfo {pages} {1} (\bibinfo {year} {2017})},\ \bibinfo {note}
  {geometry and non-adiabatic response in quantum and classical
  systems}\BibitemShut {NoStop}%
\bibitem [{\citenamefont {Landau}(1932)}]{landau}%
  \BibitemOpen
  \bibfield  {author} {\bibinfo {author} {\bibfnamefont {L.~D.}\ \bibnamefont
  {Landau}},\ }\bibfield  {title} {\bibinfo {title} {Zur theorie der
  energieubertragung ii},\ }\href@noop {} {\bibfield  {journal} {\bibinfo
  {journal} {Z. Sowjetunion}\ }\textbf {\bibinfo {volume} {2}},\ \bibinfo
  {pages} {46} (\bibinfo {year} {1932})}\BibitemShut {NoStop}%
\bibitem [{\citenamefont {Zener}(1932)}]{zener}%
  \BibitemOpen
  \bibfield  {author} {\bibinfo {author} {\bibfnamefont {C.}~\bibnamefont
  {Zener}},\ }\bibfield  {title} {\bibinfo {title} {Non-adiabatic crossing of
  energy levels},\ }\href@noop {} {\bibfield  {journal} {\bibinfo  {journal}
  {Proceedings of the Royal Society of London. Series A, Containing Papers of a
  Mathematical and Physical Character}\ }\textbf {\bibinfo {volume} {137}},\
  \bibinfo {pages} {696} (\bibinfo {year} {1932})}\BibitemShut {NoStop}%
\bibitem [{\citenamefont {Majorana}(2020)}]{majorana}%
  \BibitemOpen
  \bibfield  {author} {\bibinfo {author} {\bibfnamefont {E.}~\bibnamefont
  {Majorana}},\ }\bibfield  {title} {\bibinfo {title} {Oriented atoms in a
  variable magnetic field},\ }in\ \href@noop {} {\emph {\bibinfo {booktitle}
  {Scientific Papers of Ettore Majorana}}}\ (\bibinfo  {publisher} {Springer},\
  \bibinfo {year} {2020})\ pp.\ \bibinfo {pages} {77--88}\BibitemShut {NoStop}%
\bibitem [{\citenamefont {Stueckelberg}(1932)}]{stueckelberg1932theory}%
  \BibitemOpen
  \bibfield  {author} {\bibinfo {author} {\bibfnamefont {E.}~\bibnamefont
  {Stueckelberg}},\ }\bibfield  {title} {\bibinfo {title} {Theory of continuous
  absorption of oxygen at 1450a},\ }\href@noop {} {\bibfield  {journal}
  {\bibinfo  {journal} {Physical Review}\ }\textbf {\bibinfo {volume} {42}},\
  \bibinfo {pages} {518} (\bibinfo {year} {1932})}\BibitemShut {NoStop}%
\bibitem [{\citenamefont {Shevchenko}\ \emph {et~al.}(2010)\citenamefont
  {Shevchenko}, \citenamefont {Ashhab},\ and\ \citenamefont
  {Nori}}]{SHEVCHENKO20101}%
  \BibitemOpen
  \bibfield  {author} {\bibinfo {author} {\bibfnamefont {S.}~\bibnamefont
  {Shevchenko}}, \bibinfo {author} {\bibfnamefont {S.}~\bibnamefont {Ashhab}},\
  and\ \bibinfo {author} {\bibfnamefont {F.}~\bibnamefont {Nori}},\ }\bibfield
  {title} {\bibinfo {title} {Landau–zener–stückelberg interferometry},\
  }\href {https://doi.org/https://doi.org/10.1016/j.physrep.2010.03.002}
  {\bibfield  {journal} {\bibinfo  {journal} {Physics Reports}\ }\textbf
  {\bibinfo {volume} {492}},\ \bibinfo {pages} {1} (\bibinfo {year}
  {2010})}\BibitemShut {NoStop}%
\bibitem [{\citenamefont {Dykhne}(1962)}]{dykhne}%
  \BibitemOpen
  \bibfield  {author} {\bibinfo {author} {\bibfnamefont {A.~M.}\ \bibnamefont
  {Dykhne}},\ }\bibfield  {title} {\bibinfo {title} {Adiabatic perturbation of
  discrete spectrum states},\ }\href@noop {} {\bibfield  {journal} {\bibinfo
  {journal} {Sov. Phys. JETP}\ }\textbf {\bibinfo {volume} {14}} (\bibinfo
  {year} {1962})}\BibitemShut {NoStop}%
\bibitem [{\citenamefont {Davis}(1976)}]{davis}%
  \BibitemOpen
  \bibfield  {author} {\bibinfo {author} {\bibfnamefont {J.~P.}\ \bibnamefont
  {Davis}},\ }\bibfield  {title} {\bibinfo {title} {{Nonadiabatic transitions
  induced by a time-dependent Hamiltonian in the semiclassical/adiabatic limit:
  The two-state case}},\ }\href {https://doi.org/10.1063/1.432648} {\bibfield
  {journal} {\bibinfo  {journal} {The Journal of Chemical Physics}\ }\textbf
  {\bibinfo {volume} {64}},\ \bibinfo {pages} {3129} (\bibinfo {year}
  {1976})}\BibitemShut {NoStop}%
\bibitem [{\citenamefont {Hwang}\ and\ \citenamefont
  {Pechukas}(1977)}]{pechukas}%
  \BibitemOpen
  \bibfield  {author} {\bibinfo {author} {\bibfnamefont {J.}~\bibnamefont
  {Hwang}}\ and\ \bibinfo {author} {\bibfnamefont {P.}~\bibnamefont
  {Pechukas}},\ }\bibfield  {title} {\bibinfo {title} {{The adiabatic theorem
  in the complex plane and the semiclassical calculation of nonadiabatic
  transition amplitudes}},\ }\href {https://doi.org/10.1063/1.434630}
  {\bibfield  {journal} {\bibinfo  {journal} {The Journal of Chemical Physics}\
  }\textbf {\bibinfo {volume} {67}},\ \bibinfo {pages} {4640} (\bibinfo {year}
  {1977})}\BibitemShut {NoStop}%
\bibitem [{\citenamefont {Berry}(1990)}]{berrygeometric}%
  \BibitemOpen
  \bibfield  {author} {\bibinfo {author} {\bibfnamefont {M.~V.}\ \bibnamefont
  {Berry}},\ }\bibfield  {title} {\bibinfo {title} {{Geometric amplitude
  factors in adiabatic quantum transitions}},\ }\href
  {https://doi.org/10.1098/rspa.1990.0096} {\bibfield  {journal} {\bibinfo
  {journal} {Proceedings of the Royal Society of London. Series A: Mathematical
  and Physical Sciences}\ }\textbf {\bibinfo {volume} {430}},\ \bibinfo {pages}
  {405} (\bibinfo {year} {1990})}\BibitemShut {NoStop}%
\bibitem [{\citenamefont {Joye}\ \emph {et~al.}(1991)\citenamefont {Joye},
  \citenamefont {Kunz},\ and\ \citenamefont {Pfister}}]{joyegeometric}%
  \BibitemOpen
  \bibfield  {author} {\bibinfo {author} {\bibfnamefont {A.}~\bibnamefont
  {Joye}}, \bibinfo {author} {\bibfnamefont {H.}~\bibnamefont {Kunz}},\ and\
  \bibinfo {author} {\bibfnamefont {C.-E.}\ \bibnamefont {Pfister}},\
  }\bibfield  {title} {\bibinfo {title} {{Exponential decay and geometric
  aspect of transition probabilities in the adiabatic limit}},\ }\href
  {https://doi.org/10.1016/0003-4916(91)90297-l} {\bibfield  {journal}
  {\bibinfo  {journal} {Annals of Physics}\ }\textbf {\bibinfo {volume}
  {208}},\ \bibinfo {pages} {299} (\bibinfo {year} {1991})}\BibitemShut
  {NoStop}%
\bibitem [{\citenamefont {Joye}\ and\ \citenamefont
  {Pfister}(1999)}]{joyeexpansion}%
  \BibitemOpen
  \bibfield  {author} {\bibinfo {author} {\bibfnamefont {A.}~\bibnamefont
  {Joye}}\ and\ \bibinfo {author} {\bibfnamefont {C.~E.}\ \bibnamefont
  {Pfister}},\ }\bibfield  {title} {\bibinfo {title} {{Full asymptotic
  expansion of transition probabilities in the adiabatic limit}},\ }\href
  {https://doi.org/10.1088/0305-4470/24/4/012} {\bibfield  {journal} {\bibinfo
  {journal} {Journal of Physics A: Mathematical and General}\ }\textbf
  {\bibinfo {volume} {24}},\ \bibinfo {pages} {753} (\bibinfo {year}
  {1999})}\BibitemShut {NoStop}%
\bibitem [{\citenamefont {Aharonov}\ and\ \citenamefont
  {Bohm}(1959)}]{aharonov1959significance}%
  \BibitemOpen
  \bibfield  {author} {\bibinfo {author} {\bibfnamefont {Y.}~\bibnamefont
  {Aharonov}}\ and\ \bibinfo {author} {\bibfnamefont {D.}~\bibnamefont
  {Bohm}},\ }\bibfield  {title} {\bibinfo {title} {Significance of
  electromagnetic potentials in the quantum theory},\ }\href@noop {} {\bibfield
   {journal} {\bibinfo  {journal} {Physical Review}\ }\textbf {\bibinfo
  {volume} {115}},\ \bibinfo {pages} {485} (\bibinfo {year}
  {1959})}\BibitemShut {NoStop}%
\bibitem [{\citenamefont {Sinitsyn}\ \emph {et~al.}(2018)\citenamefont
  {Sinitsyn}, \citenamefont {Yuzbashyan}, \citenamefont {Chernyak},
  \citenamefont {Patra},\ and\ \citenamefont {Sun}}]{integrable}%
  \BibitemOpen
  \bibfield  {author} {\bibinfo {author} {\bibfnamefont {N.~A.}\ \bibnamefont
  {Sinitsyn}}, \bibinfo {author} {\bibfnamefont {E.~A.}\ \bibnamefont
  {Yuzbashyan}}, \bibinfo {author} {\bibfnamefont {V.~Y.}\ \bibnamefont
  {Chernyak}}, \bibinfo {author} {\bibfnamefont {A.}~\bibnamefont {Patra}},\
  and\ \bibinfo {author} {\bibfnamefont {C.}~\bibnamefont {Sun}},\ }\bibfield
  {title} {\bibinfo {title} {{Integrable Time-Dependent Quantum
  Hamiltonians}},\ }\href {https://doi.org/10.1103/physrevlett.120.190402}
  {\bibfield  {journal} {\bibinfo  {journal} {Physical Review Letters}\
  }\textbf {\bibinfo {volume} {120}},\ \bibinfo {pages} {190402} (\bibinfo
  {year} {2018})},\ \Eprint {https://arxiv.org/abs/1711.09945} {1711.09945}
  \BibitemShut {NoStop}%
\bibitem [{\citenamefont {Requist}\ \emph {et~al.}(2005)\citenamefont
  {Requist}, \citenamefont {Schliemann}, \citenamefont {Abanov},\ and\
  \citenamefont {Loss}}]{abanovrequist}%
  \BibitemOpen
  \bibfield  {author} {\bibinfo {author} {\bibfnamefont {R.}~\bibnamefont
  {Requist}}, \bibinfo {author} {\bibfnamefont {J.}~\bibnamefont {Schliemann}},
  \bibinfo {author} {\bibfnamefont {A.~G.}\ \bibnamefont {Abanov}},\ and\
  \bibinfo {author} {\bibfnamefont {D.}~\bibnamefont {Loss}},\ }\bibfield
  {title} {\bibinfo {title} {Double occupancy errors in quantum computing
  operations: Corrections to adiabaticity},\ }\href
  {https://doi.org/10.1103/PhysRevB.71.115315} {\bibfield  {journal} {\bibinfo
  {journal} {Phys. Rev. B}\ }\textbf {\bibinfo {volume} {71}},\ \bibinfo
  {pages} {115315} (\bibinfo {year} {2005})}\BibitemShut {NoStop}%
\bibitem [{\citenamefont {Takayoshi}\ \emph {et~al.}(2021)\citenamefont
  {Takayoshi}, \citenamefont {Wu},\ and\ \citenamefont
  {Oka}}]{takayoshi2021nonadiabatic}%
  \BibitemOpen
  \bibfield  {author} {\bibinfo {author} {\bibfnamefont {S.}~\bibnamefont
  {Takayoshi}}, \bibinfo {author} {\bibfnamefont {J.}~\bibnamefont {Wu}},\ and\
  \bibinfo {author} {\bibfnamefont {T.}~\bibnamefont {Oka}},\ }\bibfield
  {title} {\bibinfo {title} {Nonadiabatic nonlinear optics and quantum
  geometry—application to the twisted schwinger effect},\ }\href@noop {}
  {\bibfield  {journal} {\bibinfo  {journal} {SciPost Physics}\ }\textbf
  {\bibinfo {volume} {11}},\ \bibinfo {pages} {075} (\bibinfo {year}
  {2021})}\BibitemShut {NoStop}%
\bibitem [{Note1()}]{Note1}%
  \BibitemOpen
  \bibinfo {note} {The argument generalizes to the case where the Hamiltonian
  extends analytically to only a stripe around the real axis and containing the
  closes degeneracy point of the eigenstates, but the distinction does not play
  a role here.}\BibitemShut {Stop}%
\bibitem [{Note2()}]{Note2}%
  \BibitemOpen
  \bibinfo {note} {The line integrals along the curve $\gamma ^B$, $\DOTSI
  \intop \ilimits@ _0^{\tau =\gamma ^B(s)}$ are defined to be on the path: from
  zero, down the real axis to $-\infty $, then to $\gamma (-\infty )$ and then
  along $\gamma ^B$ up to $\gamma ^B(s)$. This definition (as a limit) makes
  sense provided that the Hamiltonian $h(\tau )$ smoothly approaches
  well-defined limits $h_{\pm }$ as $\protect \text {Re}[\tau ]\to \pm \infty $
  \cite {pechukas}.}\BibitemShut {Stop}%
\bibitem [{\citenamefont {Dirac}(1931)}]{dirac1931quantised}%
  \BibitemOpen
  \bibfield  {author} {\bibinfo {author} {\bibfnamefont {P.~A.~M.}\
  \bibnamefont {Dirac}},\ }\bibfield  {title} {\bibinfo {title} {Quantised
  singularities in the electromagnetic field},\ }\href@noop {} {\bibfield
  {journal} {\bibinfo  {journal} {Proceedings of the Royal Society of London.
  Series A, Containing Papers of a Mathematical and Physical Character}\
  }\textbf {\bibinfo {volume} {133}},\ \bibinfo {pages} {60} (\bibinfo {year}
  {1931})}\BibitemShut {NoStop}%
\bibitem [{\citenamefont {Nishimura}\ and\ \citenamefont
  {Takahashi}(2018)}]{countermultilz}%
  \BibitemOpen
  \bibfield  {author} {\bibinfo {author} {\bibfnamefont {K.}~\bibnamefont
  {Nishimura}}\ and\ \bibinfo {author} {\bibfnamefont {K.}~\bibnamefont
  {Takahashi}},\ }\bibfield  {title} {\bibinfo {title} {{Counterdiabatic
  Hamiltonians for multistate Landau-Zener problem}},\ }\href
  {https://doi.org/10.21468/scipostphys.5.3.029} {\bibfield  {journal}
  {\bibinfo  {journal} {SciPost Physics}\ }\textbf {\bibinfo {volume} {5}},\
  \bibinfo {pages} {029} (\bibinfo {year} {2018})},\ \Eprint
  {https://arxiv.org/abs/1805.06662} {1805.06662} \BibitemShut {NoStop}%
\bibitem [{\citenamefont {Chernyak}\ and\ \citenamefont
  {Sinitsyn}(2021)}]{multilz}%
  \BibitemOpen
  \bibfield  {author} {\bibinfo {author} {\bibfnamefont {V.~Y.}\ \bibnamefont
  {Chernyak}}\ and\ \bibinfo {author} {\bibfnamefont {N.~A.}\ \bibnamefont
  {Sinitsyn}},\ }\bibfield  {title} {\bibinfo {title} {{Integrability in the
  multistate Landau-Zener model with time-quadratic commuting operators}},\
  }\href {https://doi.org/10.1088/1751-8121/abe427} {\bibfield  {journal}
  {\bibinfo  {journal} {Journal of Physics A: Mathematical and Theoretical}\
  }\textbf {\bibinfo {volume} {54}},\ \bibinfo {pages} {115204} (\bibinfo
  {year} {2021})}\BibitemShut {NoStop}%
\bibitem [{\citenamefont {Suzuki}\ and\ \citenamefont
  {Nakazato}(2021)}]{gridlz}%
  \BibitemOpen
  \bibfield  {author} {\bibinfo {author} {\bibfnamefont {T.}~\bibnamefont
  {Suzuki}}\ and\ \bibinfo {author} {\bibfnamefont {H.}~\bibnamefont
  {Nakazato}},\ }\bibfield  {title} {\bibinfo {title} {{Generalized Adiabatic
  Impulse Approximation}},\ }\href
  {https://doi.org/10.1103/physreva.105.022211} {\bibfield  {journal} {\bibinfo
   {journal} {arXiv}\ }\textbf {\bibinfo {volume} {105}},\ \bibinfo {pages}
  {022211} (\bibinfo {year} {2021})},\ \Eprint
  {https://arxiv.org/abs/2112.04739} {2112.04739} \BibitemShut {NoStop}%
\bibitem [{\citenamefont {Malla}\ \emph {et~al.}(2021)\citenamefont {Malla},
  \citenamefont {Chernyak},\ and\ \citenamefont {Sinitsyn}}]{malla}%
  \BibitemOpen
  \bibfield  {author} {\bibinfo {author} {\bibfnamefont {R.~K.}\ \bibnamefont
  {Malla}}, \bibinfo {author} {\bibfnamefont {V.~Y.}\ \bibnamefont
  {Chernyak}},\ and\ \bibinfo {author} {\bibfnamefont {N.~A.}\ \bibnamefont
  {Sinitsyn}},\ }\bibfield  {title} {\bibinfo {title} {Nonadiabatic transitions
  in landau-zener grids: Integrability and semiclassical theory},\ }\href
  {https://doi.org/10.1103/PhysRevB.103.144301} {\bibfield  {journal} {\bibinfo
   {journal} {Phys. Rev. B}\ }\textbf {\bibinfo {volume} {103}},\ \bibinfo
  {pages} {144301} (\bibinfo {year} {2021})}\BibitemShut {NoStop}%
\bibitem [{\citenamefont {Yuzbashyan}(2018)}]{gaudin}%
  \BibitemOpen
  \bibfield  {author} {\bibinfo {author} {\bibfnamefont {E.~A.}\ \bibnamefont
  {Yuzbashyan}},\ }\bibfield  {title} {\bibinfo {title} {{Integrable
  time-dependent Hamiltonians, solvable Landau–Zener models and Gaudin
  magnets}},\ }\href {https://doi.org/10.1016/j.aop.2018.01.017} {\bibfield
  {journal} {\bibinfo  {journal} {Annals of Physics}\ }\textbf {\bibinfo
  {volume} {392}},\ \bibinfo {pages} {323} (\bibinfo {year} {2018})},\ \Eprint
  {https://arxiv.org/abs/1802.01571} {1802.01571} \BibitemShut {NoStop}%
\bibitem [{\citenamefont {Berry}\ and\ \citenamefont
  {Lim}(1993)}]{berry1993universal}%
  \BibitemOpen
  \bibfield  {author} {\bibinfo {author} {\bibfnamefont {M.}~\bibnamefont
  {Berry}}\ and\ \bibinfo {author} {\bibfnamefont {R.}~\bibnamefont {Lim}},\
  }\bibfield  {title} {\bibinfo {title} {Universal transition prefactors
  derived by superadiabatic renormalization},\ }\href@noop {} {\bibfield
  {journal} {\bibinfo  {journal} {Journal of Physics A: Mathematical and
  General}\ }\textbf {\bibinfo {volume} {26}},\ \bibinfo {pages} {4737}
  (\bibinfo {year} {1993})}\BibitemShut {NoStop}%
\end{thebibliography}%

\appendix

\section{Dynamical and geometric exponents}\label{sec:appexponent}

In evaluating the integrals in the dynamical and geometric exponents appearing in (\ref{eq:modLZ}), one can again deform the contour to lie on the imaginary axis. However, this means crossing the branch cut at $\tau^X=i$ (see figure \ref{fig:AGP}). For example, the dynamical phase becomes
\begin{align}
    &\text{Im}[\int_0^{\tau_1^*}(e_1-e_0)]=\\
    &\text{Re}[2\lim_{\varepsilon\to 0}\left(\int_0^{1-\varepsilon}-\int_{1+\varepsilon}^{\sqrt{1+a^{\frac{2}{3}}}}\right)\sqrt{1-y^2+\frac{a^2}{(1-y^2)^2}}dy],
\end{align}
where
\begin{equation}
    a=\frac{\eta\delta}{2}
\end{equation}
is small. One important consequence of the change of sign at the cut is that the integral is finite. The leading term can be easily found from the first integral at $a=0$,
\begin{equation}
    \int_0^{1}\sqrt{1-y^2}dy=\frac{\pi}{4},
\end{equation}
and just gives the LZ exponent. To get the next term in $a$, one has to carefully expand the second integral. One way is to write the integrand as
\begin{equation}
    \int_{1+\varepsilon}^{\sqrt{1+a^{\frac{2}{3}}}}\frac{\sqrt{a^{\frac{4}{3}}-a^{\frac{2}{3}}(1-y^2)+(1-y^2)^2}\sqrt{1-y^2+a^{\frac{2}{3}}}}{1-y^2}dy,
\end{equation}
which correctly separates out the pole at the lower limit and the zero at the upper limit of the integral, and expand the first square-root in $a^\frac{2}{3}$. We find
\begin{equation}
    \int_{1+\varepsilon}^{\sqrt{1+a^{\frac{2}{3}}}}\sqrt{1-y^2+a^{\frac{2}{3}}}=\frac{a}{3}+...,
\end{equation}
which gives the leading value of the dynamical phase for small $\delta$ as
\begin{equation}
    e^{-\frac{2}{\delta}\text{Im}[\int_0^{\tau_1^*}(e_1-e_0)]}=e^{-\pi/\delta+2\eta/3}.
\end{equation}

\section{Time-Dependent Integrable Quantum Systems}\label{sec:appintegrable}

We consider real symmetric Hamiltonians $H_{\mu}$ satisfying
\begin{align}
    \partial_{\mu} H_{\nu}-\partial_{\nu} H_{\mu} = 0,\label{eq:derivativesflat2}\\
    [H_{\mu},H_{\nu}]=0\label{eq:commutatorflat2}.
\end{align}
From (\ref{eq:commutatorflat2}), one can find a unitary $U$ such that 
\begin{equation}
    \tilde{H}_{\mu}=U^{\dagger}H_{\mu}U
\end{equation}
are all diagonal. Then the AGPs are
\begin{equation}
    \mathcal{A}_{\mu}=-iU\partial_{\mu}U^{\dagger},
\end{equation}
and, using that $(\partial_{\mu}U)U^{\dagger}=-U\partial_{\mu}U^{\dagger}$, we find
\begin{align}
    [\mathcal{A}_{\mu},\mathcal{A}_{\nu}]&=-(U\partial_{\mu}U^{\dagger}U\partial_{\nu}U^{\dagger}-U\partial_{\nu}U^{\dagger}U\partial_{\mu}U^{\dagger})\\
    &=\partial_{\mu}U\partial_{\nu}U^{\dagger}-\partial_{\nu}U\partial_{\mu}U^{\dagger}\\
    &=\partial_{\mu}(U\partial_{\nu}U^{\dagger})-\partial_{\nu}(U\partial_{\mu}U^{\dagger}),
\end{align}
so that the $\mathcal{A}_{\mu}$ satisfy the the flatness condition
\begin{equation}
    \partial_{\mu}\mathcal{A}_{\nu}-\partial_{\nu}\mathcal{A}_{\mu}+i[\mathcal{A}_{\mu},\mathcal{A}_{\nu}]=0.\label{eq:counterflat}
\end{equation}

Now, from (\ref{eq:derivativesflat2}),
\begin{align}
    0&=\partial_{\mu}(U\tilde{H}_{\nu}U^{\dagger})-\partial_{\nu}(U\tilde{H}_{\mu}U^{\dagger})\\
    &=U(\partial_{\mu} \tilde{H}_{\nu}-\partial_{\nu} \tilde{H}_{\mu})U^{\dagger}+[H_{\nu},U\partial_{\mu}U^{\dagger}]-[H_{\mu},U\partial_{\nu}U^{\dagger}].
\end{align}
Conjugating by $U$ we get
\begin{equation}
    (\partial_{\mu} \tilde{H}_{\nu}-\partial_{\nu} \tilde{H}_{\mu})+([\tilde{H}_{\nu},(\partial_{\mu}U^{\dagger})U]-[\tilde{H}_{\mu},(\partial_{\nu}U^{\dagger})U])=0.
\end{equation}
Since the $\tilde{H}_{\mu}$ are diagonal, the contribution from the first round brackets is diagonal and that from the second is off-diagonal. They separately give
\begin{align}
    \partial_{\mu}E_{\nu,n}-\partial_{\nu}E_{\mu,n}&=0,\\
    [H_{\mu},\mathcal{A}_{\nu}]-[H_{\nu},\mathcal{A}_{\mu}]&=0.\label{eq:commuteperp}
\end{align}
Finally, using (\ref{eq:derivativesflat2}),(\ref{eq:commutatorflat2}), (\ref{eq:counterflat}) and (\ref{eq:commuteperp}), we find that
\begin{equation}
    \partial_{\mu}(H_{\nu}+\mathcal{A}_{\nu})-\partial_{\nu}(H_{\mu}+\mathcal{A}_{\mu})+i[H_{\mu}+\mathcal{A}_{\mu},H_{\nu}+\mathcal{A}_{\nu}]=0,
\end{equation}
so that the Hamiltonians corrected by the AGP terms still satisfy the flatness condition. In other words, the adiabatic gauge potential is consistent with time-dependent quantum integrability.

\end{document}